# Diamagnetic Effects, Spin Dependent Fermi Surfaces, and the Giant Magnetoresistance in Metallic Multilayers

W. Tavera and G. G. Cabrera

*Instituto de Física "Gleb Wataghin",*
*Universidade Estadual de Campinas (UNICAMP),*
*C. P. 6165, Campinas, SP 13083-970, Brazil*

(Received)

## Abstract

We study the role of diamagnetic effects on the transport properties of metallic magnetic multilayers to elucidate whether they can explain the *Giant Magnetoresistance* (GMR) effect observed in those systems. Realistic Fermi surface topologies in layered ferromagnets are taken into account, with the possibilities of different types of orbits depending on the electron spin. Both configurations, with ferromagnetic and anti-ferromagnetic couplings between magnetic layers, are considered and the transmission coefficient for scattering at the interface boundary is modelled to include magnetic and roughness contributions. We assume that scattering processes conserve the electron spin, due to large spin diffusion lengths in multilayer samples. Scattering from the spacer mixes different orbit topologies in a way similar to magnetic *breakdown* phenomena. For antiferromagnetic coupling, majority and minority spins are interchanged from one magnetic layer to the next. Cyclotron orbits are also traveled in opposite directions, producing a compensation-like effect that yields a huge GMR, particularly for closed orbits. For open orbits, one may get the *inverse* magnetoresistance effect along particular directions.

The Giant Magnetoresistance (GMR) effect discovered in magnetic multilayers[1] is known to have many potential applications, including recording materials[2] and new promising electronic devices based on spin effects.[3,4] This field of magnetism, that now involves surface physics, metallic films, magnetic small particles and nanostructures, is currently very active in both, basic and applied research. A typical multilayer system consists of ferromagnetic films (usually iron or cobalt), separated by thin non-magnetic metal layers. The GMR effect appears when the ferromagnetic films are coupled through the spacer in an antiparallel configuration. The application of an external magnetic field yields a negative magnetoresistance, which is typically of the order of 50~60% for the Current In-Plane (CIP) set up. The effect is attributed to the existence of spin dependent electronic scattering[5,6], in such a way, that transport properties can be described in terms of a *two current* model, with two spin channels[7]. Further developments include modelling of the interface roughness and quantum effects in very clean samples[8]. Within the semi-classical approach, most of the calculations in the literature integrate the Boltzmann equation with different boundary conditions, following the original proposal by Fuchs and Sondheimer for thin films[5,6]. Diamagnetic effects of the electron orbits have been neglected when compared to the interface scattering, which is consider as dominant in most contributions. We would like to argue in this paper that this is not a sensible approximation. In a ferromagnetic metal, like iron or cobalt, there is a huge internal magnetic field, resulting in very different Fermi surfaces for each spin[9–11]. As a consequence of the large spin splitting, we may have different electron occupations at the Fermi level for each spin band, and depending on the direction of the internal field, we may vary the orbit topology in a different manner for each spin. This is the case of iron, where the minority Fermi sheet supports an open orbit along the $<001>$ direction. In contrast, the majority Fermi surface displays a multiply-connected open orbit in the $<110>$ direction[9]. A recent band calculation for ferromagnetic fcc cobalt[11] shows a variety of situations: both spins present closed orbits for the (001) plane; on the other hand, for the (210) plane, the majority-spin orbits are all closed while the minority Fermi surface supports open orbits. Most of the scattering mechanisms at the spacer, which is paramagnetic or non-magnetic, conserve the spin (no spin-flip). With the present state of the art in fabricating multilayers, the spin diffusion length for spin polarized carriers are of the order of the sample dimensions[4]. This means that, whether transmitted or reflected back at the interfaces between layers, the carriers remains in the same spin Fermi sheet. For the anti-parallel configuration, the internal magnetic field is reversed when going from one magnetic layer to the next, with the corresponding interchange between the majority and minority spin Fermi surfaces. Similar effects has been considered by Cabrera and Falicov[13], to study transport near magnetic domain walls. The same concepts lead to the following scenario: **i)** a different availability of states for each spin at the Fermi level, producing a *spin valve* effect when the electronic current flows through the multilayer structure; **ii)** due to the strong internal field, diamagnetic effects are important. Since the Fermi surfaces are different for the two spin states, scattering from the spacer couples electron orbits of different forms and/or topologies at both sides of the interface, in a way similar to magnetic breakdown phenomena[12]. As usual, diamagnetic effects are important when $\chi \equiv \omega_c \tau \gg 1$, with $\omega_c$ being the cyclotron frequency and $\tau$ the bulk mean collision time.

In this paper, we propose a simple model that incorporates the above affects, with the presence of the sample walls as an essential ingredient that may affect the connectivity of the cyclotronic orbits. If the mean collision time for scattering from the interface is much smaller than the cyclotron period $T_c$, electrons at the Fermi surface have no time to explore the orbit topology, even if the condition $\chi \gg 1$ is satisfied. The interesting case corresponds to the situation when both time scales are of the same order, which implies mixed orbit topologies. We note that at high magnetic fields, the qualitative magnetoresistance behavior does not depend on details of the Fermi surface, but only on its topology[14]. Simple spherical shapes are then used in our numerical calculation to model closed and open orbits. The sample geometry adopted in our model, is that of a three-layer *sandwich*, with wide ferromagnetic layers separated by a thin non-magnetic spacer. In a simplified picture, the spacer is considered as an interface between magnetic layers, whose roughness is modelled through a potential barrier $V_0$, independent of the spin. In addition, due to the exchange splitting $V_1$, the electrons are subjected to different magnetic potentials, depending on their spin. We have calculated transport properties for the two magnetic configurations, the one with anti-parallel coupling between magnetic layers (APC), and the other one with parallel alignment (PC), and measured their difference. We assume that the *sandwich* is repeated periodically, to generate the multilayer structure.

To calculate the conductivity tensor, we use the Chambers' path integral approach, which is particularly suited to treat the contribution of entangled networks of electron orbits[15]. Within this scheme, the conductivity is given by

$$\sigma = \frac{e^2}{8\pi^3\hbar} \int_{FS} \frac{\mathbf{v}_\lambda(\mathbf{k})\mathbf{\Lambda}_{\mathbf{k}\lambda}}{|\mathbf{v}_\lambda(\mathbf{k})|} dS \ , \qquad (1)$$

with $\mathbf{v}_\lambda(\mathbf{k})$ being the velocity vector for an electron on the Fermi surface, and

$$\mathbf{\Lambda}_{\mathbf{k}\lambda}(\mathbf{r}) = \int_{-\infty}^{t_0(\mathbf{k})} \mathbf{v}_{\mathbf{k}\lambda}(t) exp(\frac{t - t_0(\mathbf{k})}{\tau}) dt, \qquad (2)$$

being the effective mean free path for wave vector $\mathbf{k}$ and spin $\lambda$. The above formulae have been established using the Chambers kinetic approach[15], which is equivalent to solving the Boltzmann equation, taking into account terms to all orders in the magnetic field. As usual, the time dependent $\mathbf{v}_{\mathbf{k}\lambda}(t) = (1/\hbar)\nabla_k \epsilon_\lambda(\mathbf{k})$ in (2), represents the velocity of an electron which is in the state $\mathbf{k}$ at time $t_0(\mathbf{k})$, and is obtained solving the equation:

$$\hbar \dot{\mathbf{k}} = -\frac{|e|}{c}\mathbf{v_k} \times \mathbf{B} \ . \qquad (3)$$

The integration of (3) yields a $\pi/2$ rotation between the orbits in real and $k$-space, apart from a scale factor that depends on the magnetic field. We choose the $x$-axis as perpendicular to the interfaces, with the internal magnetic field being $\mathbf{B} = B\hat{z}$ or $\mathbf{B} = -B\hat{z}$, depending on the kind of magnetic coupling between ferromagnetic layers. The cyclotron orbits are then realized in the $xy$-plane, and we are just calculating components of the relevant tensors projected on that plane. We take a complete network of orbits and divide it into $n$ suitable pieces named $\alpha, \beta, ...$, to describe the proper Fermi surface topology. Further, we label the state $\mathbf{k}$ of an electron by a piece-number and a time $t$. Electrons moving on a particular piece orbit $\alpha$ are in the states $(t, \alpha)$, with $0 \leq t \leq t_\alpha$. We assume that at some points on each orbit piece, due to the interface scattering, there is a probability for an electron to be transmitted to a different piece orbit or to be reflected back, remaining in the same Fermi surface sheet.

For simplicity, we have considered the same time $t_\alpha$ for all the pieces (*isochronous pieces*), thus $t_\alpha$ is the elapsed time for electrons, moving on the $\alpha$-piece, between two scattering processes from the interface. Different networks, with different connectivities, are generated depending on the time scales of the problem, *i.e.* depending on the parameter $\eta = t_\alpha/T_c$, where $T_c$ is the cyclotron period.

With this notation, we may write relation (2) in terms of a discrete set of quantities $\{\mathbf{\Lambda}_{(\alpha)}\}$, in the form

$$\mathbf{\Lambda}(t,\alpha) = \mathbf{\Lambda}_{(\alpha)} e^{-t/\tau} + e^{-t/\tau} \int_0^t \mathbf{v}(t',\alpha) e^{t'/\tau} dt' \tag{4}$$

where $\alpha = 1,...,n$. The $\mathbf{\Lambda}_{(\alpha)}$ terms take all the information about the past history of electrons on the network before reaching the $\alpha$−piece at time $t = 0$. In the limit case of total reflection from the interface, electrons remain in pre-defined orbits for each spin state. On the other hand, in the limit of total transmission, the electron moves in a complex, but coherent, network of orbits mixing the Fermi topology of the magnetic layers. So, one gets a kind of artificial *'breakdown'* interface-driven mechanism between this two limits. Further, we include all the information about the connectivity among the pieces, defining a suitable square *probability matrix* $\mathbf{M}$, which is calculated for both spin orientations, *up* ($\uparrow$) and *down* ($\downarrow$), and for the two magnetic configurations, APC and PC. The component $M_{\alpha\beta}$ of the matrix above, represent the probability that an electron which goes into piece $\alpha$ will come from piece $\beta$. The latter takes into account conservation laws for scattering processes at the interface.

Defining $[\mathbf{\Lambda}]$ as being a column matrix whose components are the vectors $\mathbf{\Lambda}_{(\alpha)}$ ($\alpha = 1,...,n$), and integrating (4) backwards in time, we can write the following matrix equation for $\mathbf{\Lambda}_{(\alpha)}$:

$$[\mathbf{\Lambda}] = \mathbf{M}\{\mathbf{E}[\mathbf{\Lambda}] + [\mathbf{V}]\}, \tag{5}$$

where $[\mathbf{V}]$ (a column matrix) and $\mathbf{E}$ (a diagonal square matrix) are spin and configuration dependent. The components of $[\mathbf{V}]$ and $\mathbf{E}$ have the form

$$\mathbf{V}_\alpha = e^{-t_\alpha/\tau} \int_0^{t_\alpha} \mathbf{v}(t',\alpha) e^{t'/\tau} dt', \quad E_{\alpha\beta} = e^{-t_\alpha/\tau} \delta_{\alpha\beta} .$$

Inserting the solution of (5) in (4), and then in (1), we obtain the conductivity tensor. Finally, by tensor inversion of the latter, we obtain the resistivity. The magnetoresistance ($\Delta\rho/\rho$) is defined here as

$$\Delta\rho/\rho = (\rho_{\uparrow\downarrow} - \rho_{\uparrow\uparrow})/(\rho_{\uparrow\downarrow} + \rho_{\uparrow\uparrow}) , \tag{6}$$

where $\rho_{\uparrow\downarrow}$ and $\rho_{\uparrow\uparrow}$ are the resistivities of the sample for APC and PC, respectively. We note that the magnetoresistance considered here is due to the internal magnetic field, which for a given sample is a constant. To vary $\chi \equiv \omega_c \tau$ in relation (6), implies measurements in several samples with different purities and internal fields (neglecting variations with temperature).

We consider several topologies and connectivities induced by different values of $\eta$. The limit cases of high ($T \to 1$) and low ($T \to 0$) transmissions, are shown in Fig. (1). The examples include open orbits for majority (minority) spin and closed ones for minority (majority), and closed orbits for the two spins. In all calculations we assumed a constant and isotropic effective mass of carriers; furthermore, we consider the mean collision time $\tau$, as being the same across the sample. The spin dependent behavior is then entirely due to Fermi surface effects. We vary the height and the width of the barrier at the spacer, assuming values given in the literature for some metallic multilayers. The magnetic potential, depending on the exchange splitting, is assumed to be of the order of $eV$, as in the case of iron[6].

Some symmetry considerations are in order. In the APC configuration, when going from one magnetic layer to the next, majority and minority spins are interchanged. Thus, electrons with spin ↑ and ↓ have the same density and topology in the limit cases depicted in Fig. 1. Antiparallel magnetic fields in neighboring layers drive the electrons in orbits with different curvature, and one may get compensated-like orbits in some cases. This effect is clear when we consider closed orbits for the two spin states (see Fig. 1(c)), but it can happen in other cases in which open orbits break down in identical closed orbits, in the high transmission limit (the case shown in Fig. 1(b) is an example of this effect). Compensated-like orbits highly enhance the magnetoresistance in both directions $x$ and $y$ in a nearly isotropic fashion. Compensation between the two spin states for the APC case, yields a quadratic increase of the resistivity with the field. On the other hand, for the PC configuration, all closed orbits are electron-like, leading to saturation of the magnetoresistance. The net effect for $\Delta\rho/\rho$ is a huge magnetoresistance effect, which is shown in Fig.2 as a function of the dimensionless parameter $\chi = \omega_c \tau$.

We have also made calculations that include open orbits, with open orbit orientations perpendicular (Fig. 1(a)) and parallel (Fig. 1(b)) to the interface. Figure 3 shows some particular cases calculated. All the curves are plots of $\Delta\rho/\rho$ as a function of $\chi$. The magnetoresistance is anisotropic when an open orbit is present. In same cases we may get $\Delta\rho/\rho < 0$ values, *i.e. inverse* magnetoresistance effect[16] (see left side of Fig.3).

In summary, diamagnetic effects are important due to high internal fields in ferromagnets. We have shown that compensation-like effects in the APC configuration yield high values for the magnetoresistance, when the scattering time $t_\alpha$ is equal or longer than the cyclotron period $T_c$. In general, when $\eta \gg 1$, electrons have enough time to probe the Fermi surface topology, before being scattered at the interface.

On the other hand, under particular conditions that lead to open orbits, one may get the *inverse magnetoresistance* effect. The latter phenomenon depends critically on the *isochronous* assumption for the orbit pieces that are connected by scattering at the interface. For a general distribution of scattering times, electron orbits will trace random walk trajectories, with the corresponding loss of coherence.

Scattering from the interface between magnetic layers produces, in general, a mixing of orbits as in *'breakdown'* phenomena, suggesting a mechanism for the GMR based exclusively on Fermi surface effects. Our results indicate that the so called Kohler's rule is not valid, since there are two relevant parameters, $\chi$ and $\eta$, in the theory. The authors are currently extending the present calculation to include a finite spin-flip relaxation time, *i.e.* a finite spin diffusion length.[17]

The authors acknowledge partial financial support from *Conselho de Desenvolvimento Científico e Tecnológico* (CNPq, Brazil) through projects # 143009/95-8 and # 301221/77-4.

FIGURES

FIG. 1. Connectivities and orbit topologies considered in our calculation, in the limits $T \to 0$ and $T \to 1$, for the transmission coefficient at the interface. Cases (a) and (b) correspond to $\eta = 1/2$, and (c) to $\eta = 1$. We depict both magnetic arrangements, with the APC and PC configurations. In (a), orbits for majority and minority spins that are closed at $T \to 0$, breakdown to open ones in the direction perpendicular to the interface, when $T \to 1$. In (b), we have, at $T \to 0$, closed orbits for minority spin and open ones for majority. They breakdown, at $T \to 1$, to different topologies depending on the magnetic configuration: for the APC setup, we get compensated-like closed orbits; for the PC one, we get an open orbit for minority spin plus closed ones for majority. Case (c) is the same as (a) at $T \to 0$, but now the different value of $\eta$ ($\eta = 1$) implies a different connectivity at $T \to 1$. In this latter we get exclusively closed orbits, which are compensated only for the APC case. The thick line for the PC setup means that orbits are retraced.

FIG. 2. Magnetoresistance as a function of $\omega_c \tau$ for case (c) of Fig.1, near the high transmission regime ($T \approx 1$). The parameters are indicated in the figure (mass of the carriers $m$, Fermi energy $E_F$, height of the interface barrier $V_0$, and width of the spacer $d$). The spin splitting $V_1$ is also indicated in the inset of the upper figure. The magnetoresistance is nearly isotropic, as shown in the figure for the $x$ and $y$ directions. Note that the huge GMR effect is due to compensation in the APC setup, and is essentially independent of the splitting and the transmission coefficient at the barrier.

FIG. 3. Magnetoresistance as a function of $\omega_c \tau$ for the cases (a) and (b) of Fig.1, at fixed exchange splitting and for several values of the interface barrier $V_0$, which are shown in the upper right figure. Other parameters are the same as in the previous figure. Both cases are anisotropic, due to the presence of open orbits. Case (a) displays the *inverse* magnetoresistance effect along the $y$ direction, for values of the parameters near the case $T \approx 1$.

|  | T→0 | T→1 | T→0 | T→1 | T→0 | T→1 | SPIN |  |
|---|---|---|---|---|---|---|---|---|
|  |  |  |  |  |  |  | ↓ | A |
|  |  |  |  |  |  |  |  | P |
|  |  |  |  |  |  |  | ↑ | C |
|  |  |  |  |  |  |  | ↓ | P |
|  |  |  |  |  |  |  | ↑ | C |
|  | (a) |  | (b) |  | (c) |  |  |  |

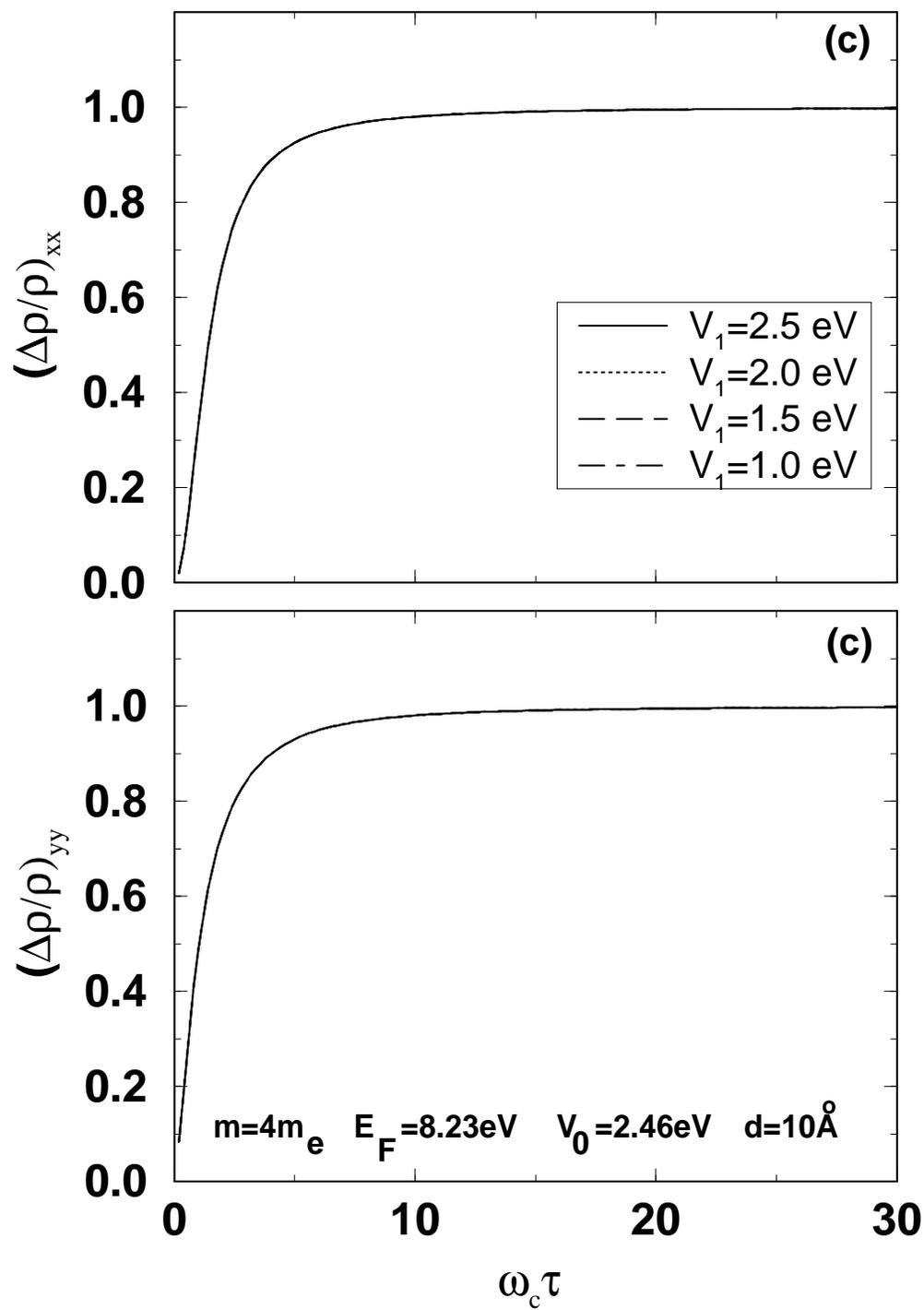

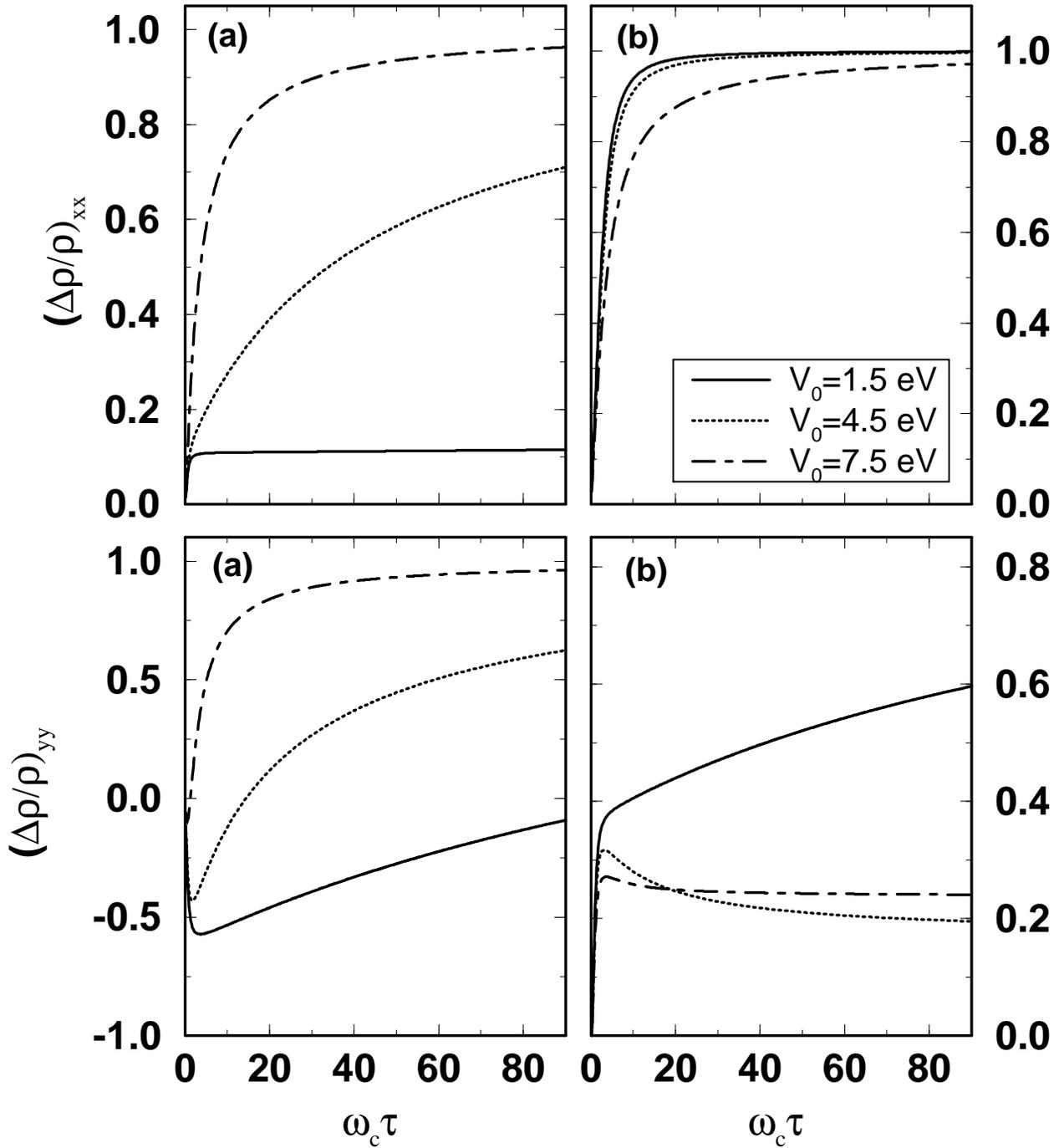